\documentclass[preprintnumbers,amsmath,amssymb,floatfix,10pt,prd,onecolumn,
superscriptaddress,nofootinbib]{revtex4}

\usepackage[parfill]{parskip}    

\usepackage{graphicx}
\usepackage{pdfpages}
\usepackage{amssymb}
\usepackage{epstopdf}
\usepackage{color}
\usepackage{float}
\usepackage{amsmath}
\usepackage{tikz-cd}
\usepackage[colorlinks=true,citecolor=blue,urlcolor=magenta,breaklinks]{hyperref}

\begin{document}

\title{Gravitational--gauge vector interaction in the Ho\v{r}ava--Lifshitz framework   }

\author{Alvaro Restuccia}
\email{alvaro.restuccia@uantof.cl}
\affiliation{Departamento de F\'isica, Facultad de Ciencias Básicas, Universidad de Antofagasta, Casilla 170, Antofagasta, Chile.}

\author{Francisco Tello-Ortiz }
\email{francisco.tello@ua.cl}
\affiliation{Departamento de F\'isica, Facultad de Ciencias Básicas, Universidad de Antofagasta, Casilla 170, Antofagasta, Chile.}

\begin{abstract}
An anisotropic model describing gravity--vector gauge coupling at all energy scales is presented. The starting point is the 4+1 dimensional non--projectable Ho\v{r}ava--Lifshitz gravity theory subject to a geometrical restriction. Renormalizability arguments require all possible interactions in the potential up to terms with $z=4$ spatial derivatives on the geometrical tensor fields: the Riemann and Weyl tensors. The latter being necessary on a 4+1 dimensional formulation. The dimensional reduction to 3+1 dimensions give rise to a model invariant under {foliation--preserving diffeomorphisms} (FDiff) and $U(1)$ symmetry groups. The reduced theory on the {kinetic conformal} (KC) point ($\lambda =1/3$), propagates the same spectrum of the Einstein--Maxwell theory. Moreover, at low energies, on the IR point $\alpha=0$, $\beta=1$, its field equations are exactly the Einstein--Maxwell ones in a particular gauge condition. The Minkowski ground state is stable provided several restrictions on the coupling parameters are satisfied, they are explicitly obtained. The quantum propagators of the physical degrees of freedom are obtained and after an analysis of the first and second class constraints the renormalizability by power counting is proved, provided that the aforementioned restrictions on the coupling parameters are satisfied.

\end{abstract}
\maketitle

\section{Introduction}\label{sec1}

Horava--Lifshitz proposal describes an interesting framework to approach quantum gravity. It is based on an anisotropic scaling realized by introducing a dynamical critical exponent $z$ \cite{r1,lif}. The final version of the formulation includes the interaction terms introduced in \cite{r3}. The theory is formulated on a 3+1 foliation with a privileged temporal direction described in terms of the well--known ADM variables \cite{r2}, the 3--dimensional metric $\gamma_{ij}$ of the spacelike leaves, the lapse $N$ and shift $N_{i}$ functions. Besides, the symmetry group is the foliation--preserving diffeomorphisms (FDiff), where the spatial diffeormorphisms generators depend on space and time, whilst the time generator depends solely in the time coordinate.   

Due to its anisotropic scaling, Horava's theory admits the introduction of terms with high spatial derivatives of the metric $\gamma_{ij}$ and the function $N$ up to $2z$ order, this renders the theory to be power--counting renormalizable and unitary provided several restrictions on its coupling parameters are satisfied, otherwise the above property is not valid.\footnote{Contrary to what happens in General Relativity (GR from now on), where high order covariant derivatives terms lead to a renormalizable but a non--unitary theory \cite{r7}. This is so because, those high order covariant derivative terms contain both high spatial and temporal derivatives, where high temporal derivatives introduce ghost fields.} \cite{r1,r3}. The inclusion of $N$ in the potential of the theory, can only be done in the non--projectable version of the theory \cite{r1,r3}, since in the projectable case, the lapse function $N$ depends only on time variable \cite{r1}. Therefore, the projectable and non--projectable models propagate different spectra. What is more, the non--projectable version contains two separate cases, namely, when the dimensionless coupling parameter $\lambda$, different from $1/d$, is arbitrary and when it takes its critical value $\lambda=1/d$ (being $d$ the spatial dimension of spacelike leaves of the foliation) \cite{r6,r19}. In the former case, the theory propagates the gravitational modes, \i.e., the transverse--traceless tensorial modes, and a scalar field\footnote{The theoretical study of the non--projectable theory with this extra scalar mode was analyzed in several contributions and, few of them are \cite{r4,r5,rr5}. Moreover, some phenomenological aspects of Ho\v{r}ava's theory were analyzed in \cite{bara}. In this concern, it should be noted that the presence of the scalar degree of freedom imposes strong restrictions to the coupling constants of the theory.}. In \cite{r8}, to eliminate the scalar field, an extension of the symmetry group, including the $U(1)$ gauge symmetry was considered. With the same goal, within the framework of the so--called mimetic Ho\v{r}ava--gravity \cite{l1,l2}, a dynamically broken diffeomorphism invariance by means of the coupling of a non--standard scalar field to curvature, was introduced. In contrast with the mentioned approaches, instead of extending the group of symmetries or introducing new fields, the so--called conformal kinetic point formulation (KC point hereinafter) \cite{r6}, eliminates the additional degree of freedom by fixing the coupling $\lambda$ to its critical value. In contrast with the non--projectable version with arbitrary $\lambda$, the non--projectable case at the KC point, propagates only the transverse traceless tensorial modes as GR. Moreover, as shown in \cite{r24444} these tensorial modes obey the same quadrupole radiation formula as the relativistic graviton in GR. In the KC point, the theory has an additional second--class constraint with respect to the version with $\lambda\neq 1/d$, the conjugate momenta to the spatial metric is traceless. Precisely, this second--class constraints and their preservation in time are consistently solved to eliminate the additional mode and find out the Lagrange multipliers associated with this constraint.    

 Ho\v{r}ava--Lifshitz gravity has been widely related with some other theories presenting Lorentz symmetry breaking. For example, at second order in derivatives (truncated Ho\v{r}ava's potential theory), both Einstein--aether \cite{r9} and the non--projectable Ho\v{r}ava--Lifshitz gravity theories coincide \cite{r5,r10,r11}. In a different framework, taking into account geometrized Newtonian gravitation or Newton--Cartan (NC) theory, the projectable version corresponds to the dynamic NC geometry without torsion while the non--projectable version corresponds to the dynamic NC geometry with torsion without twist \cite{r12}. Furthermore, Ho\v{r}ava--Lifshitz gravity has been analyzed on the cosmological arena \cite{r23,r24,r25,r26}, the standard model domain \cite{r27} and the four--fermion Gross--Neveu like models \cite{r28}, to name a few. In the quantum gravity context, the Hamiltonian formulation of the non--projectable version of
Ho\v{r}ava's proposal, has been analyzed thoroughly \cite{r13,r14,r16,r15,new1,new2,new3,new4}. Moreover, the features of the Ho\v{r}ava--Lifshitz model on the KC point at the classical and quantum level have been studied in \cite{r6,r18}. In this direction, beyond the usual Hamiltonian setting to study quantum properties, the quantization of the theory has been tested following the so--called background--covariant formalism \cite{blas2,blas3} and the stochastic quantization procedure \cite{chin1}. See also \cite{barvinsky,by} for the quantization in 2+1 dimensions. 

To further support the feasibility of Ho\v{r}ava--Lifshitz gravity, some theoretical studies concerning the coupling matter fields to this anisotropic gravity theory have been carried out. This fact is quite relevant at both classical at quantum level, since it is well--known that when matter fields are coupled to GR, divergences at loops order become worst \cite{r32}. In this direction in \cite{r31}, applying the minimal coupling matter procedure, relativistic matter fields in the Ho\v{r}ava action were introduced. Nevertheless, in such a case the resulting theory, is not longer power--counting renormalizable. This is so because, quadratic divergences in the matter sector percolate to the gravity sector. In order to cure the above issue, an alternative scheme to couple matter fields\footnote{See \cite{ian} for coupling matter fields in Ho\v{r}ava--Lifshitz gravity, including Feynman diagram analysis at tree level. } to Ho\v{r}ava--Lifshitz gravity, was proposed in \cite{r35,r36}. In their study, the authors proposed a new class of interacting terms, namely the so--called \emph{mixed derivative terms}. However, despite the final theory keeps power--counting renormalizability, ghost fields arise (states with negative kinetic energy). Therefore, the unitarity of the theory is violated. On the other hand, a different approach has been considered to couple matter fields to the non--projectable version of the Ho\v{r}ava--Lifshitz gravity. Specifically in \cite{r29}, at low energy level ($z=1$) a gauge vector field was incorporated via dimensional reduction \emph{a la Kaluza--Klein} from 4+1 to 3+1 dimensions. Following this approach in \cite{r30}, at the KC point, it was proved that at low energy, where only the $z=1$ interaction terms were considered, the field equations evaluated at the space parameter values $\{\alpha;\beta\}=\{0;1\}$, exactly agree with Einstein--Maxwell field equations under a particular gauge. Interestingly, this 3+1 theory possess a spherically symmetric solution of its field equations (for generic $\alpha$) describing a charged throat, being  asymptotically flat space on one side and an essential singularity on the other one \cite{wh1}. Besides, this manifold on the asymptotically flat side coincides with the well--known Reissner--Nordstr\"{o}m solution of Einstein--Maxwell theory, when expanded in terms of $1/r$ \cite{wh2}.

Whether in the case of pure anisotropic gravity or including matter fields as mentioned before, a necessary question is the phenomenological verification of Horava's theory, that is, how close is it to GR? (for a completed revision of GR experimental support see \cite{will}). In this regard, taking into account the recent detection of gravitational waves
GW170817 and gamma rays burts GRB170817A \cite{feno1,feno2}, it was shown that both gravitational and electromagnetic waves propagation speeds, match within one part in $10^{-15}$ \cite{feno3}. In \cite{bara,feno5,feno6} restrictions on the low energy coupling constants $\{\alpha;\beta\}$ were determined, providing $\{\beta\sim 1, \alpha \sim 0\}$ for the pure gravitational anisotropic sector. These numerical values are in the same admissible range as the relativistic data, according to low energy observations. Following the theoretical proposal in \cite{r29,r30}, where it was shown that at low energies both, the gravitational and gauge vector waves propagate at the same speed $\sqrt{\beta}$, the authors in \cite{feno7} employing observational data from GRB170817A in a cosmological FLRW background showed that  $|1-\sqrt{\beta}|<(10^{-19}-10^{-18})$.

The main aim of this article, is to provide a complete description at all energy scales, of the pure anisotropic gravity--gauge vector field coupling in the non--projectable Ho\v{r}ava--Lifshitz framework. To do so, we use as a mathematical device a 4+1--dimensional Ho\v{r}ava--Lifshitz model, with a foliated manifold structure $\mathbb{M}_{4}\times \mathbb{R}$ (where $\mathbb{M}_{4}$ is a Riemannian manifold which locally has the structure $\mathbb{M}_{3}\times \mathbb{S}^{1}$), subject to a geometrical restriction. The reduced 3+1--dimensional theory propagates only the gravitational transverse traceless tensorial degrees of freedom and the emergent transverse gauge vector field ones. Since, the quantum renormalizability requires all possible terms in the potential compatible with the symmetries of the theory up to $z=4$ terms, the contribution of the Weyl tensor to the potential becomes then necessary. Therefore, relevant terms constructed from the 4--dimensional Weyl tensor contribute to the gravitational and gauge vector sectors at all orders \i.e., $z=2,3,4 $.     

Once the dimensional reduction has been performed, we analyze the resulting 3+1--dimensional theory on the KC point ($\lambda=1/3$ in this case). It directly eliminates the natural scalar degree of freedom of the Ho\v{r}ava--Lifshitz gravity, ending with a theory invariant under 3+1 FDiff and $U(1)$ gauge group\footnote{This is the same action obtained in \cite{r30} (see Eq. (22) in that article).}, which propagates two transverse--traceless tensorial modes and two transverse vector modes, as Einstein--Maxwell theory. Then, the consistency of the Hamiltonian and its constraints at UV scale, is analyzed under a perturbative scheme, determining the superficial degree of divergence\footnote{To obtain the superficial degree of divergence, we have employed the same techniques developed in \cite{r37,r38, an, an1,an2,weinberg}.}, showing that the power--counting renormalization is guaranteed provided several restrictions on the coupling parameters are satisfied.

The outlook of the article is as follows: In Sec. \ref{sec2} the Hamiltonian formulation and its the dimensional reduction are presented. Sec. \ref{sec3} is devoted to the stability analysis of the ground state and Sec. \ref{sec4} presents the power--counting renormalization procedure. Finally, Sec. \ref{sec5} concludes the research. 

\section{The Model}\label{sec2}

In this section we formulate a Ho\v{r}ava--Lifshitz model in a 5--dimensional foliated manifold $\mathbb{M}_{4}\times \mathbb{R}$, where $\mathbb{M}_{4}$ is a Riemannian manifold which locally has the structure $\mathbb{M}_{3}\times \mathbb{S}^{1}$. Globally it will have a more general structure as we will discuss shortly. $\mathbb{M}_{3}$ is asymptotically diffeomorphic to $\mathbb{R}^{3}\backslash \mathbb{B}$ when $\mathbb{B}$ is a 3--dimensional open ball of large enough radius.

The fields describing the gravitational model are the 4--dimensional metric on $\mathbb{M}_{4}$, $g_{\mu\nu}$, $\mu= (i,4)$ and $i=1, 2, 3$, the lapse $N$ and shift $N_{\mu}$ which describe the foliated manifold $\mathbb{M}_{4}\times \mathbb{R}$.

The diffeomorphisms on $\mathbb{M}_{3}\times \mathbb{S}^{1}$ are not the same as the diffeomorphisms on a 4--dimensional manifold asymptotically diffeomorphic to $\mathbb{R}^{4}\backslash \mathbb{B}$, with $\mathbb{B}$ a 4--dimensional open ball of large enough radius. In fact, to analyze the diffeomorphisms on $\mathbb{S}^{1}$, we describe $\mathbb{S}^{1}$ as $\mathbb{R}\backslash \mathbb{Z}$, that is $\mathbb{R}$ modulo translations $T$
    \begin{equation}
y \rightarrow y+a,
\end{equation}
where $y$ is a coordinate on $\mathbb{R}$ and $a$ the length of $\mathbb{S}^{1}$. We consider now the subgroup of diffeomorphisms on $\mathbb{R}$, Diff, which commutes with the translations $T$
 \begin{eqnarray}
y &\stackrel{\text { Diff }}{\longrightarrow}&f(y)\stackrel{T}{\longrightarrow} f(y)+a, \\
y &\stackrel{T}{\longrightarrow}&y+a\stackrel{\text { Diff }}{\longrightarrow} f(y+a).
\end{eqnarray}
Hence $f(y+a)=f(y)+a$. We denote it Diff$^{\mathbb{Z}}$.

If we describe $\mathbb{S}^{1}$ as a circle of length $a$ on the complex plane $\mathbb{C}$, the action of 
Diff$^{\mathbb{Z}}$ on it becomes
\begin{equation}
    \text{Exp}\left(\frac{2\pi i}{a}y\right)\rightarrow \text{Exp}\left(\frac{2\pi i}{a} f(y)\right),
\end{equation}
which is a well defined diffeomorphism $\mathbb{S}^{1}\rightarrow \mathbb{S}^{1}$. Since 
\begin{equation}
    \text{Exp}\left(\frac{2\pi i}{a}f(y+a)\right)\rightarrow \text{Exp}\left(\frac{2\pi i}{a} f(y)\right).
\end{equation}

We then have the exact sequence 
\begin{equation}
    T\longrightarrow \text{Diff}^{\mathbb{Z}} \twoheadrightarrow \text{Diff}\left(\mathbb{S}^{1}\right).
\end{equation}
In fact, the translation $y\longrightarrow y+a$ becomes the kernel in the transformation Diff$^{\mathbb{{Z}}}\longrightarrow$ Diff$(\mathbb{S}^{1})$
since 
\begin{equation}
    \text{Exp}\left(\frac{2\pi i}{a}y\right)\stackrel{T}{\longrightarrow} \text{Exp}\left(\left(\frac{2\pi i}{a}\right)(y+a)\right)=\text{Exp}\left(\frac{2\pi i}{a}y\right),
\end{equation}
is the identity in Diff$(\mathbb{S}^{1})$. The map $\twoheadrightarrow$ is a surjection. The group of diffeomorphisms acting on $\mathbb{M}_{3}\times \mathbb{S}^{1}$ is then given by
\begin{eqnarray}\label{eq8new}
x^{i} &\longrightarrow& x^{\prime i}= x^{\prime i} (x, y, t), \\ \label{eq9new}
y &\longrightarrow& y^{\prime} = y^{\prime} (x, y, t), \\ \label{eq10new}
t^{\prime} &\longrightarrow& t^{\prime}= f(t),
\end{eqnarray}
where 
\begin{eqnarray}
x^{\prime i} (x, y+a, t)&=&x^{\prime i} (x, y, t)+a, \\
y^{\prime } (x, y+a, t)&=&y^{\prime} (x, y, t)+a.
\end{eqnarray}
Under (\ref{eq8new}) and (\ref{eq9new}), $g_{\mu\nu}$ transforms as a tensor and under (\ref{eq10new}) as a scalar field. The lapse $N$ transforms under (\ref{eq8new}) and (\ref{eq9new}) as a scalar field and under (\ref{eq10new}) as a density, that is,
\begin{equation}
    N^{\prime}(x^{\prime}, y^{\prime}, t^{\prime}) dt^{\prime} = N(x, y, t) dt.
\end{equation}
The shift $N_{\mu}$ transforms under (\ref{eq8new}) and (\ref{eq9new}) as a Lagrange multiplier and under (\ref{eq10new}) as a density.

We notice that $g_{\mu\nu}(x, y, t)$, $N(x, y, t)$ and $N_{\mu}(x, y, t)$ are periodic functions on $y$, that is
\begin{equation}\label{neweq14}
    \varphi(x, y+a, t)= \varphi(x, y, t).
    \end{equation}

On this geometrical framework, and following Ho\v{r}ava's proposal, we consider the anisotropic scaling
\begin{equation}
    x, y \sim b, \quad t \sim b^{z},
\end{equation}
and take $z=$ space--like dimension $=4$. It ensures that $k_{5}$ below is dimensionless. Then, the Ho\v{r}ava--Lifshitz action is given by 
\begin{equation}\label{eq15new}
    S_{\text{H--L}}=\frac{1}{k_{5}}\int_{\mathbb{M}_{4}\times \mathbb{R}} dt\,dy\,d^{3}xN\sqrt{g}\left[K_{\mu\nu}K^{\mu\nu}-\lambda K^{2}+\beta \textbf{R}+\alpha a_{\mu}a^{\mu} -\mathcal{V}(g_{\mu\nu}, a_{\rho})\right],
\end{equation}
where $K_{\mu\nu}$ is the extrinsic curvature tensor, $K$ its trace, $\textbf{R}$ the 4--dimensional scalar curvature, $a_{\mu}=\partial_{\mu} N/N$ and $\mathcal{V}(g_{\mu\nu}, a_{\rho})$ is the high space--like
derivative potential, a scalar field. $\lambda$, $\beta$ and $\alpha$ are coupling parameters, being $\lambda$ dimensionless. In this work we consider the 5--dimensional construction as a mathematical device to obtain the Ho\v{r}ava--Lifshitz interaction terms in 3+1 dimensions. It provides the complete expression of the high order in space--like derivatives potential which in the reduced 3+1 dimensional theory describe the high order interaction terms of the gravitational, electromagnetic and scalar fields. The model is a power counting renormalizable theory, provided the coupling parameters satisfy some consistency relations which we will obtain in the following sections.

We can always choose the coordinates in $\mathbb{M}_{3}\times \mathbb{S}^{1}$ such that the metric component associated to $\mathbb{S}^{1}$ is fixed. We take
\begin{equation}\label{gauge}
    g_{44}=1.
\end{equation}
Although (\ref{gauge}) is a coordinate choice, it is not an off--shell restriction of (\ref{eq15new}).

We may also consider the action
\begin{equation}\label{eq17new}
    S=S_{\text{H--L}}+\int_{\mathbb{M}_{4}\times \mathbb{R}} dt\,dy\,d^{3}x\,\Lambda (g_{44}-1),
\end{equation}
where (\ref{gauge}) is now imposed as an off--shell condition. 
Both actions are power counting renormalizable theories and under dimensional reduction will give rise to different consistent gravity theories, interacting with a gauge vector field and in some cases with scalar fields. {In this concern, the restriction (\ref{gauge}) eliminates in a consistent way the scalar field introduced by the dimensional reduction procedure. This leads, considering the action (\ref{eq17new}), to a pure anisotropic gravity--gauge vector field coupling as desired (see below for further details).}\\

Now, it is convenient to rewrite the metric $g_{\mu\nu}$ as
\begin{equation}\label{eq18new}
g_{\mu\nu} =\begin{pmatrix}
\gamma_{ij}+ \phi A_{i}A_{j} & \quad  \phi A_{j}\\
\phi A_{i} & \phi
\end{pmatrix},
\end{equation}
where $\gamma_{ij}$ is a 3--dimensional Riemannian metric, {$A_{i}$ a vector field and $\phi$ a scalar field under FDiff.} The inverse metric becomes
\begin{equation}\label{eq16}
g^{\mu\nu}=\begin{pmatrix}
\gamma^{ij} &  -A^{j}\\
-A^{i} & \quad \frac{1}{\phi}+A_{k}A^{k}
\end{pmatrix},
\end{equation}
where $\gamma^{ij}$ are the components of the inverse of $\gamma_{ij}$ and $A^{i}=\gamma^{ij}A_{j}$.

We now consider a dimensional reduction procedure. In order to implement it, we must give to $\mathbb{M}_{4}$, which is locally $\mathbb{M}_{3}\times \mathbb{S}^{1}$, the geometrical structure of a fiber bundle with total space $E$, base manifold $\mathbb{M}_{3}$, fiber $\mathbb{S}^{1}$ and structure group the diffeomorphisms consistent with the fiber bundle diagram  

\[\begin{tikzcd}
          E   \arrow[rr, "\text{Diff}"] 
        \arrow[d, "\Pi",swap]  
        & & E' \arrow[d, "\Pi'"] \\
        \mathbb{M}_{3} \arrow[rr, "\text{Diff}"] 
        & & \mathbb{M}'_{3}\\
\end{tikzcd}\]

The group of diffeomorphisms reduce then to the subgroup 
\begin{equation}
t^{\prime}=t^{\prime}(t), \quad x^{\prime i}=x^{\prime i} (x^{j}, t), \quad y^{\prime}= y^{\prime} (x^{j}, y, t),
\end{equation}
where 
\begin{equation}\label{eq21new}
 y^{\prime} (x^{j}, y+a, t)= y^{\prime} (x^{j}, y, t) + a.
\end{equation}
The diffeomorphisms in $\mathbb{R}$ which commute with the translations.
This is the most general group of diffeomorphisms preserving the fiber bundle structure of $\mathbb{M}_{4}$.

If we impose $g_{44}=\phi=1$, then (\ref{eq21new}) reduces to
\begin{eqnarray}\nonumber
x^{\prime i} &=& x^{\prime i} (x^{j}, t), \\ \label{eq22new}
y^{\prime}&=& y+\xi(x^{j}, t).
\end{eqnarray}
A different argument is the following, from (\ref{eq21new}) if we expand the components of the metric $\gamma_{ij}$, $A_{i}$ and $N$ and $N_{\rho}$ as Fourier expansions
\begin{equation}
    \varphi=\varphi_{n}(x)\text{Exp}\left(\frac{2\pi\, i}{a}ny\right),
\end{equation}
(since they are periodic functions on $y$) and consider the subgroups of (\ref{eq21new}) preserving the Fourier basis, we also obtain the subgroup of diffeomorphisms (\ref{eq22new}), without using (\ref{neweq14}).

The massive modes have mass
\begin{equation}
    m=\frac{n}{r},
\end{equation}
where $r=a/2\pi$ is the radius of $\mathbb{S}^{1}$. Since, in this work we consider the 4+1 dimensional construction as a mathematical device to obtain the H--L interaction terms from a pure gravity theory, we perform the dimensional reduction by taking $r\rightarrow 0$. The mass terms then decouple from the massless ones. 

Starting from $S_{\text{H--L}}$ given by (\ref{eq15new}), we can perform the dimensional reduction procedure and end up with the 3+1 action given in \cite{r29}. On the other hand, starting with (\ref{eq17new}) and, following also a dimensional reduction we will construct a different theory which propagates solely the transverse traceless--tensorial and transverse vectorial degrees of freedom as the Einstein--Maxwell theory. We will include all the high order derivative terms in the potential according to the Ho\v{r}ava--Lifshitz  proposal.
In what follows we will consider the dimensional reduction of the action $S$ (\ref{eq17new}).

We can impose $g_{44}=1$ directly into the action, since this restriction has been introduced into the action through a Lagrange multiplier. It is an on--shell as well as off--shell restriction.

The expression for $\mathcal{V}(\gamma_{ij}, A_{k}, a_{j})$ involves not only the 3--dimensional Ricci tensor as in the 3+1 construction, but also the Weyl tensor which is nontrivial in a 4--dimensional Riemannian manifold. So we have up to quadratic terms on the tensor fields
\begin{eqnarray}\label{eq50}
    \mathcal{V}^{(z=1)}&=&  -\beta \textbf{R}-\alpha a_{\mu}a^{\mu}, \\ \label{eq51}
     \mathcal{V}^{(z=2)}&=&-\beta_{2}\textbf{R}^{2}-\beta_{1}\textbf{R}_{\mu\nu}\textbf{R}^{\mu\nu}-\alpha_{2}\nabla^{\mu}a^{\nu}\nabla_{\mu}a_{\nu}- \alpha_{1}\textbf{R}\nabla_{\mu}a^{\mu}-\kappa_{1} \textbf{C}_{\mu\nu\rho\sigma}\textbf{C}^{\mu\nu\rho\sigma}, \\ \label{eq52}
    \mathcal{V}^{(z=3)}&=& -\beta_{4}\nabla_{\mu}\textbf{R}\nabla^{\mu}\textbf{R}-\beta_{3}\nabla_{\mu}\textbf{R}_{\nu\rho}\nabla^{\mu}\textbf{R}^{\nu\rho}-\alpha_{4} \nabla^{2} a_{\mu}\nabla^{2} a^{\mu}-\alpha_{3}\nabla^{2}\textbf{R} \nabla_{\mu}a^{\mu}-\kappa_{2}\nabla_{\theta}\textbf{C}_{\mu\nu\rho\sigma}\nabla^{\theta}\textbf{C}^{\mu\nu\rho\sigma}, \\ \label{eq53}
    \mathcal{V}^{(z=4)}&=& -\beta_{6}\nabla^{2} \textbf{R}\nabla^{2} \textbf{R}-\beta_{5}\nabla^{2}\textbf{R}_{\mu\nu}\nabla^{2} \textbf{R}^{\mu\nu}-\alpha_{6}\nabla^{2} \nabla_{\mu}a_{\nu}\nabla^{2} \nabla^{\mu}a^{\nu}-\alpha_{5}\nabla^{4}
    \textbf{R}\nabla_{\mu}a^{\mu}-\kappa_{3}\nabla_{\omega}\nabla_{\theta}\textbf{C}_{\mu\nu\rho\sigma}\nabla^{\omega}\nabla^{\theta}\textbf{C}^{\mu\nu\rho\sigma},
\end{eqnarray}
where $\beta's$, $\alpha's$ and $\kappa's$ are coupling constants and $\nabla^{2}\equiv \nabla_{\mu}\nabla^{\mu}$. We have only explicitly introduced the independent quadratic terms on the Riemann and Weyl tensors, since we will deal generically with high--order polynomial terms when necessary ({For further details concerning the contribution of the high order polynomic terms, see appendix \ref{appen}}). According to the Ho\v{r}ava--Lifshitz approach \cite{r1}, one has to introduce up to order eight spatial derivative terms in the potential,  since the theory is formulated in 4+1 dimensions. It will then lead to a power counting renormalizable theory. The explicit expression of the potential after performing the dimensional reduction procedure is then given by
\begin{align}\label{eq54}
{\mathcal{V}}^{(z=1)}=&-{\beta{R}}+\frac{\beta}{4}F_{ij}F^{ij}-{\alpha a_{i}a^{i}}, \\ \label{eq55}
{\mathcal{V}}^{(z=2)}=&-\left({\beta_{1}}+{2\kappa_{1}}\right){R}_{ij}{R}^{ij}-\left(\frac{\beta_{1}}{2}+{\kappa_{1}}\right)\nabla_{l}F^{\ l}_{i}\nabla_{m}F^{im}-\left({\beta_{2}}+{\frac{2}{3}\kappa_{1}}\right){R}^{2}-{\alpha_{1}{R}\nabla_{i}a^{i}}-{\alpha_{2}\nabla^{i}a^{j}\nabla_{i}a_{j}}, \\ \label{eq56}
{\mathcal{V}}^{(z=3)}=&-\left({\beta_{3}}+{2\kappa_{2}}\right)\nabla_{i}{R}_{jk}\nabla^{i}{R}^{jk}-\left(\frac{\beta_{3}}{2}+{\kappa_{2}}\right)\nabla_{k}\nabla_{l}F^{\ l}_{i}\nabla^{k}\nabla_{m}F^{im}-\left({\beta_{4}}+{\frac{2}{3}\kappa_{2}}\right)\nabla_{i}{R}\nabla^{i}{R}-\alpha_{3}\nabla^{2}{R}\nabla_{i}a^{i} \nonumber \\ &-\alpha_{4}\nabla^{2}a_{i}\nabla^{2}a^{i},
\\ \label{eq57}
{\mathcal{{V}}}^{(z=4)}=& -\left({\beta_{5}}+{2\kappa_{3}}\right)\nabla^{2}{R}_{ij}\nabla^{2}{R}^{ij}-\left(\frac{\beta_{5}}{2}+{\kappa_{3}}\right)\nabla^{2}\nabla_{l}F^{\ l}_{i}\nabla^{2}\nabla_{m}F^{im}-\left({\beta_{6}}+{\frac{2}{3}\kappa_{3}}\right)\nabla^{2}{R}\nabla^{2}{R}-\alpha_{5}\nabla^{4}{R}\nabla_{i}a^{i} \nonumber \\ & -\alpha_{6}\nabla^{2}\nabla_{i}a_{j}\nabla^{2}\nabla^{i}a^{j}.
\end{align}
where we have used 
\begin{equation}\label{eq58}
\textbf{R}={R}-\frac{1}{4}F_{ij}F^{ij},
\end{equation}
and for the Weyl tensor one has \cite{r42,r43,r44}
\begin{align}\label{eq59}
\textbf{C}^{ijkl}=& {C}^{ijkl}+2\left(\gamma^{i[k} c^{l] j}-\gamma^{j[k} c^{l] i}\right)
+\frac{1}{4}\left(F^{il} F^{kj}-F^{ik} F^{lj}+2 F^{ij} F^{kl}\right)
+\frac{3}{2}\left(\gamma^{i[k} T^{l] j}-\gamma^{j[k} T^{l] i}\right), \\ \label{eq60}
\textbf{C}^{4ijk}=&\frac{1}{2} \nabla^{i} F^{jk}+\frac{1}{2} \gamma^{i[j} \nabla_{m} F^{k] m}-A_{m} C^{mijk}, \\ \label{eq61}
\textbf{C}^{4k4l}=&-2c^{kl}-A_{i}A_{j}\left(\gamma^{ij}c^{lk}-\gamma^{il}c^{jk}-\gamma^{kl}c^{li}+\gamma^{kl}c^{ij}\right),
\end{align}
where
\begin{align} \label{eq62}
c^{ij}=&\frac{1}{2}\left[{R}^{ij}-\frac{1}{3} \gamma^{ij} {R}+F^{ik} F^{\ j}_{k}-\frac{1}{3} \gamma^{ij} F^{lk} F_{kl}\right], \\ \label{eq63}
T^{ij}=&F^{ik} F^{\ j}_{k}-\frac{1}{4} \gamma^{ij} F^{kl} F_{lk}.
\end{align}
{In the above expressions, the object $F_{ij}$ is defined as follows
\begin{equation}
    F_{ij}\equiv \partial_{i}A_{j}-\partial_{j}A_{i}.
\end{equation}}

The reduced potential (\ref{eq54})--(\ref{eq57}) contains up to 8 spatial derivatives and includes all possible quadratic tensorial contributions for both, the gravitational and the gauge vector sector. 

{At this stage, it is worth mentioning that, the expressions (\ref{eq50})--(\ref{eq53}) include only the independent quadratic terms contributing to the propagator of the physical degrees of freedom. Nevertheless, in general the potential of the theory contains terms such as $\textbf{R}^{3}$, $\textbf{R}\textbf{R}_{\mu\nu}\textbf{R}^{\mu\nu}$, $a_{\mu}a_{\nu}\textbf{R}^{\mu\nu}$, $(a_{\mu}a^{\mu})^{2}$ and so on. These terms represent pure interacting terms ones and do not contribute to the propagator. Therefore, after dimensional reduction, all terms beyond the quadratic order from both sectors (the gravitational and gauge vector) coming from expressions (\ref{eq50})--(\ref{eq53}), are not included in the expressions (\ref{eq54})--(\ref{eq57}) of the reduced potential. However, all high order terms will be  generically included  in the  renormalizability argument. What is more, 
terms of the form $\textbf{R}\nabla_{\mu}\nabla_{\nu}\textbf{R}^{\mu\nu}$ are equivalent to $\nabla_{\mu}\textbf{R}\nabla^{\mu}\textbf{R}$. To give this result, an integration by parts and the usage of the contracted Bianchi's identify $\nabla_{\mu}\textbf{R}^{\mu\nu}=\frac{1}{2}\nabla^{\nu}\textbf{R}$ where employed. There are also boundary terms such as $\nabla^{2}\textbf{R}$ which do not contribute to the action at the quadratic level. So, in general the procedure to detect them and to remove them from the above list, is by employing integration by parts and some tensor identities. }

Finally, we impose the dimensional reduction condition $\partial/\partial y=0$. We end up with the following 3+1 Lagrangian density 
\begin{equation}\label{actionn}
    S(\gamma_{ij}, A_{k}, N, N_{i}, N_{4})=\frac{1}{k_{5}}\int dt\, d^{3}x\, N\sqrt{g}\left[K_{\mu\nu}K^{\mu\nu}-\lambda K^{2}+\beta R+\alpha a_{i}a^{i}+\mathcal{V}(\gamma_{ij}, A_{k}, a_{j})\right],
\end{equation}
where we have used $g_{44}=1$ and will now decompose all terms in its 3+1 components.

We remark that the component of the 4--dimensional extrinsic curvature associated to the $y$ component $K_{44}$, satisfies $K_{44}=0$. It implies the following relation 
\begin{equation}\label{eq41new}
    K_{44}=g_{4\mu}g_{4\nu}K^{\mu\nu}=A_{l}A_{m}K^{lm}+2A_{l}K^{l4}+K^{44}=0,
\end{equation}
which we will use shortly. In the following analysis we take $1/k_{5}=1$.

We now proceed to construct the canonical Hamiltonian formulation associated to (\ref{actionn}). The conjugate momentum associated to $\gamma_{ij}$ is 
\begin{equation}\label{eq42new}
    p^{ij}=\sqrt{\gamma}\left(K^{ij}-\lambda K\gamma^{ij}\right)
\end{equation}
and the conjugate momentum associated to $A_{i}$ is
\begin{equation}\label{eq43new}
    p^{i}=2\sqrt{\gamma}\left(K^{i4}-\lambda K g^{i4}\right)+2p^{ij}A_{j}.
\end{equation}
It follows the relation 
\begin{equation}\label{eq44new}
\gamma_{ij}\frac{p^{ij}}{\sqrt{\gamma}}=(1-3\lambda)K, \quad K=g^{\mu\nu}K_{\mu\nu}.
\end{equation}
We will consider in this work the case $\lambda=1/3$ since we are interested in pure gravity--gauge vector formulation without scalar fields. Consequently we have a primary constraint
\begin{equation}
    \gamma_{ij}p^{ij}=0.
\end{equation}
The case $\lambda\neq1/3$, involves the propagation of a scalar field not present in the case $\lambda=1/3$.

After some calculations we obtain for $\lambda=1/3$
\begin{equation}
    K_{\mu\nu}K^{\mu\nu}-\lambda K^{2}=\frac{p^{ij}}{\sqrt{\gamma}}\frac{p^{lm}}{\sqrt{\gamma}}\gamma_{il}\gamma_{jm}+\frac{1}{2}\frac{p^{i}}{\sqrt{\gamma}}\frac{p^{j}}{\sqrt{\gamma}}\gamma_{ij}=\frac{p^{ij}}{\sqrt{\gamma}}\frac{p_{ij}}{\sqrt{\gamma}}+\frac{1}{2}\frac{p^{i}}{\sqrt{\gamma}}\frac{p_{i}}{\sqrt{\gamma}}.
\end{equation}
Besides, we have
\begin{equation}
    p^{ij}\dot{\gamma}_{ij}+p^{i}\dot{A}_{i}=2N\sqrt{\gamma}\left(K_{\mu\nu}K^{\mu\nu}-\lambda K^{2}\right).
\end{equation}
Consequently, the Hamiltonian density becomes
\begin{equation}\label{new12}
    \mathcal{H}=p^{ij}\dot{\gamma}_{ij}+p^{i}\dot{A}_{i}-\mathcal{L}=N\sqrt{\gamma}\left[\frac{p^{ij}}{\sqrt{\gamma}}\frac{p_{ij}}{\sqrt{\gamma}}+\frac{1}{2}\frac{p^{i}}{\sqrt{\gamma}}\frac{p_{i}}{\sqrt{\gamma}}-\beta {R}+\frac{\beta}{4}F_{ij}F^{ij}-\alpha a_{i}a^{i}-\mathcal{V}(\gamma_{ij}, A_{k}, a_{j})\right]-\Lambda H-\Lambda_{j}H^{j}-\sigma P_{N}-\mu P,
\end{equation}
where the quadratic contributions to $\mathcal{V}(\gamma_{ij}, A_{k}, a_{j})$ are given in 
(\ref{eq55})--(\ref{eq57}), 
\begin{eqnarray}\label{gauss1}
H&\equiv&\partial_{i}p^{i}=0, \\ \label{mome}
H^{j}&\equiv& 2\nabla_{i}p^{ij}+p^{i}\gamma^{jk}F_{ik}=0,
\end{eqnarray}
and the Lagrange multipliers $\Lambda\equiv N_{4}$, $\Lambda_{j}\equiv N_{i}-A_{i}N_{4}$. $P_{N}$ is the conjugate momentum associated to $N$ and $P\equiv \gamma_{ij}p^{ij}$. Moreover, the Hamiltonian (\ref{new12}) is subject to the second class constraints coming from preservation in time of $P_{N}$ and $P$
\begin{eqnarray}\label{eq45}
\mathcal{{H}}\equiv\frac{1}{\sqrt{\gamma}} \left(p^{ij}p_{ij}+\frac{p^{i}p_{i}}{2}\right)+\sqrt{\gamma} \mathcal{{U}}=0,  \\ \label{eq46}
\mathcal{{C}} \equiv \frac{3}{2\sqrt{\gamma}} \left(p^{ij}p_{ij}+\frac{p^{i}p_{i}}{6}\right)-\sqrt{\gamma} \mathcal{{W}}=0,
\end{eqnarray}
where at $z=1$ the explicit form of ${\mathcal{U}}$ and ${\mathcal{W}}$ in Eqs. (\ref{eq45}) and (\ref{eq46}) are given by
\begin{equation}\label{eq11113}
{\mathcal{U}}^{(z=1)}=-\beta R +\frac{\beta}{4}  F_{ij}F^{ij} +\alpha a_{i}a^{i}
+2\alpha\nabla_{i}a^{i},    
\end{equation}
\begin{equation}\label{eq11112}
{\mathcal{W}}^{(z=1)}=-\frac{1}{2}\beta R-\frac{1}{8}\beta F^{lm}F_{lm}-\left(\frac{\alpha}{2}-2\beta\right)a^{k}a_{k}  +2\beta\nabla^{l}a_{l}, 
\end{equation}
respectively. In  general we get for $\mathcal{U}$ and $\mathcal{W}$ the following expressions
\begin{equation}
\mathcal{U} \equiv \frac{1}{\sqrt{g}} \frac{\delta}{\delta N} \int d^{3} y \sqrt{g} N \mathcal{V}=\mathcal{V}+\frac{1}{N} \sum_{r=1}(-1)^{r} \nabla_{i_{1} \cdots i_{r}}\left(N \frac{\partial \mathcal{V}}{\partial\left(\nabla_{i_{r} \cdots i_{2}} a_{i_{1}}\right)}\right)
\end{equation}
and,
\begin{equation}
    \mathcal{W} \equiv g_{i j} \mathcal{W}^{i j}, \quad \mathcal{W}^{i j} \equiv \frac{1}{\sqrt{g} N} \frac{\delta}{\delta g_{i j}} \int d^{3} y \sqrt{g} N \mathcal{V},
\end{equation}
where $\nabla_{ij\ldots k}$ stands for $\nabla_{i}\nabla_{j}\ldots \nabla_{k}$.

It is worth mentioning that, if we consider the low energy model, that is, we consider only the z=1terms in the potential, it exactly agrees with the Hamiltonian density in \cite{r30}, which was obtained from a different approach. Moreover, the field equations when $\alpha=0$ and $\beta=1$ are exactly the Einstein--Maxwell field equations.

The ground state solution is $\gamma_{ij}=\delta_{ij}$, $A_{i}=0$, $N=1$, $N_{i}=0$ and $N_{4}=0$. A necessary condition for consistency and for a perturbative quantum analysis is its classical stability. We show in the next section the stability of the ground state provided some restrictions to the coupling constants are satisfied. 

\section{The stability of the ground state}\label{sec3}

In this section we show the classical stability of the ground state 
\begin{equation}
    \gamma_{ij}=\delta_{ij}, \quad A_{i}=0, \quad N=1, \quad \Lambda=0, \quad \Lambda_{i}=0.
\end{equation}
We assume the asymptotic conditions \cite{r40}
\begin{equation}
g_{ij}-\delta_{ij}=\mathcal{O}(1 / r), \quad \partial_{k} g_{ij}=\mathcal{O}\left(1 / r^{2}\right), \quad {p^{ij}}=\mathcal{O}\left(1 / r^{2}\right), \quad N-1=\mathcal{O}(1 / r), \quad \nabla_{i} N=\mathcal{O}\left(1 / r^{2}\right),
\end{equation}
consequently
\begin{equation}
K_{ij}=\mathcal{O}\left(1 / r^{2}\right), \quad R=\mathcal{O}\left(1 / r^{3}\right), \quad a_{i}=\mathcal{O}\left(1 / r^{2}\right).
\end{equation}

Under these conditions we perform a perturbative analysis around the ground state. We consider 
\begin{equation}\label{eq64}
\begin{array}{l}
\gamma_{ij}=\delta_{ij}+\epsilon h_{i j}, \quad p^{i j}=\epsilon \vartheta^{i j} \quad
\Lambda_{i}=\epsilon n_{i}, \quad \Lambda=\epsilon n_{4}, \quad N=1+\epsilon n,
\end{array}
\end{equation}
while for the vector field $A_{i}$ and its conjugate momentum $p^{i}$ one has 
\begin{equation}\label{eq65}
A_{i}=\epsilon \xi_{i}, \quad p^{i}=\epsilon\chi^{i}.  
\end{equation}
We will use the T+L ADM decomposition \cite{r2} for the metric and its conjugate momenta, 
\begin{equation}\label{eq66}
h_{ij}=h^{TT}_{ij}+\frac{1}{2}\left(\delta_{ij}-\frac{\partial_{i}\partial_{j}}{\partial^{2}}\right)h^{T}+\partial_{i}h^{L}_{j}+\partial_{j}h^{L}_{i}.
\end{equation}
We will work under the gauge fixing condition 
\begin{equation}\label{eq67}
\partial_{i}h_{ij}=0,   
\end{equation}
which implies that the longitudinal part $h^{L}_{j}=0$. It also follows from the first class constraints,
\begin{equation}
    \vartheta^{Lj}=0
\end{equation}
and from the second class constraints $P\equiv\gamma_{ij}{p^{ij}}=0$
\begin{equation}
    \vartheta^{T}=0.
\end{equation}
From the second class constraints (\ref{eq45})--(\ref{eq46}) we obtain 
\begin{equation}\label{eq76}
\mathbb{M}\phi=0.    
\end{equation}
where $\phi$ and $\mathbb{M}$ are defined as 
\begin{equation}\label{eq73}
\phi=\left(\begin{array}{c}
h^{T} \\
n
\end{array}\right), \quad \mathbb{M}=\left(\begin{array}{ll}
\mathbb{D}_{1} & \mathbb{D}_{2} \\
\mathbb{D}_{2} & \mathbb{D}_{3}
\end{array}\right),
\end{equation}
where
\begin{align}\label{eq74}
\mathbb{D}_{1} \equiv \frac{1}{8}\left[-\left(3\hat{\beta}_{5}+8\hat{\beta}_{6}\right)\partial^{8}+\left(3\hat{\beta}_{3}+8\hat{\beta}_{4}\right)\partial^{6}-\left(3\hat{\beta}_{1}+8\hat{\beta}_{2}\right)\partial^{4}+\beta\partial^{2}\right], \\ \label{eq75}
\mathbb{D}_{2} \equiv \frac{1}{2}\left[\alpha_{5}\partial^{8}+\alpha_{3}\partial^{6}+\alpha_{1}\partial^{4}+\beta\partial^{2}\right], \quad \mathbb{D}_{3} \equiv -\alpha_{6}\partial^{8}+\alpha_{4}\partial^{6}-\alpha_{2}\partial^{4}+\alpha\partial^{2},
\end{align}
{where we have defined
\begin{align}\label{eq72}
\hat{\beta}_{1}\equiv \beta_{1}+2\kappa_{1}, \quad  \hat{\beta}_{2}\equiv \beta_{2}+\frac{2}{3}\kappa_{1},  \quad
\hat{\beta}_{3}\equiv \beta_{3}+2\kappa_{2}, \quad \hat{\beta}_{4}\equiv \beta_{4}+\frac{2}{3}\kappa_{2}, \quad
\hat{\beta}_{5}\equiv \beta_{5}+2\kappa_{3}, \quad \hat{\beta}_{6}\equiv \beta_{6}+\frac{2}{3}\kappa_{3}.
\end{align}}

 $\mathbb{M}$ is an elliptical operator of maximal order, if and only if
\begin{equation}\label{eq80}
K\equiv\frac{1}{8}\bigg[\alpha_{6}\left(8\hat{\beta}_{6}+3\hat{\beta}_{5}\right)-2\alpha^{2}_{5}\bigg]\neq 0.  
\end{equation}
This is an important assumption, otherwise the model is not renormalizable by power counting. {To further clarify this point, the coupled equations (\ref{eq76}) can be diagonalized. In fact, using $\mathbb{D}_{i}\mathbb{D}_{j}=\mathbb{D}_{j}\mathbb{D}_{i}$,  $i,j=1,2,3$, we obtain  
\begin{equation}
\left(\begin{array}{ll}
\mathbb{D}_{3}\mathbb{D}_{1}-\mathbb{D}_{2}\mathbb{D}_{2} & ~~~~~~~~~0 \\
~~~~~~~~~0 & \mathbb{D}_{3}\mathbb{D}_{1}-\mathbb{D}_{2}\mathbb{D}_{3}
\end{array}\right)\phi=0.
\end{equation}    
The symbol of the operator $\mathbb{D}_{3}\mathbb{D}_{1}-\mathbb{D}_{2}\mathbb{D}_{2}$ is 
\begin{equation}
    \hat{S}=K\left(\psi\cdot\psi\right)^{4},
\end{equation}
with $K$ given by (\ref{eq80}). If $\psi\neq 0$, $\hat{S}\neq 0$ if and only if $K\neq 0$. Consequently $\mathbb{D}_{3}\mathbb{D}_{1}-\mathbb{D}_{2}\mathbb{D}_{2}$ is an elliptic operator if and only if $K \neq 0$. Moreover, $K \neq 0$ ensures that the system (67) is of maximal order.} So, under this assumption, standard arguments on elliptic operators imply 
\begin{equation}\label{eq84}
h^{T}=n=0.    
\end{equation}
Besides, we can impose on the gauge vector the gauge fixing conditions 
\begin{equation}\label{eq92}
 \partial_{i}\xi_{i}=0.   
\end{equation}
Also from (\ref{gauss1}) we obtain $\chi^{Li}=0$.

After completing the analysis on the constraints and gauge fixing conditions, we are left with the following physical degrees of freedom 
\begin{equation}
    \{ h^{TT}_{ij}; \vartheta^{TTij}; \xi^{T}_{i}; \chi^{Ti}\}.
\end{equation}
Finally, the Hamiltonian evaluated on the constrained sub--manifold and on the gauge fixing conditions, becomes 
\begin{equation}\label{eq95}
H_{\text{RED}}=\int d^{3}x\bigg[2\kappa \vartheta^{TT}_{ij}\vartheta^{TT}_{ij}+\frac{1}{4}h^{TT}_{ij}\mathbb{V}h^{TT}_{ij}+\kappa \chi^{T}_{i}\chi^{T}_{i}+\frac{1}{2}\xi^{T}_{i}\mathbb{V}\xi^{T}_{i}\bigg],    
\end{equation}
where the selfadjoint operator $\mathbb{V}$ is given by
\begin{equation}\label{eq96}
\mathbb{V}=-\beta\partial^{2}-\hat{\beta}_{1}\partial^{4}+\hat{\beta}_{3}\partial^{6}-\hat{\beta}_{5}\partial^{8}.
\end{equation}
 We notice that the same operator $\mathbb{V}$ appears on the gravitational and gauge vector potential. This is a consequence of the Kaluza--Klein approach we have followed.

The requirement of stability imposes further restrictions on the coupling parameters. In fact, $\mathbb{V}$ must be a strictly positive operator. The dominant term at low energies is $-\beta\partial^{2}$, hence we require $\beta>0$ (which is in agreement with the experimental data). On the other side at high energies the dominant term is $-\hat{\beta}_{5}\partial^{8}$, hence we require $\hat{\beta}_{5}<0$. We now determine the complete requirements on the coupling parameters in order to have a positive operator $\mathbb{V}$.

\subsection{Strictly positive $H_{\text{RED}}$ }

We introduce the Fourier transform $\tilde{\mathbb{V}}$ of the differential operator $\mathbb{V}$. The elliptic operator $\mathbb{V}$ is positive if its spectrum is positive. We notice that $\mathbb{V}$ is positive if and only if $\tilde{\mathbb{V}}$ is positive. In fact, let us suppose that $\tilde{\mathbb{V}}$ is negative for some positive value $k^{2}$. Then, since $\hat{\beta}_{5}<0$ and $\tilde{\mathbb{V}}(0)=0$, $\tilde{\mathbb{V}}$ has a minimum $-M$, $M>0$. Under this condition, we now prove that the continuous interval $[-M,+\infty)$ belongs to the spectrum of the differential operator $\mathbb{V}$.

We consider the equation 
\begin{equation}
(\mathbb{V}+m)w(x)=G(x),    
\end{equation}
and its Fourier transform. If $-m$ does not belong to the interval $[-M,+\infty)$, then
\begin{equation}
\tilde{w}=\frac{\tilde{G}}{\tilde{\mathbb{V}}+m}.
\end{equation}
Consequently, if $\tilde{G}(k)\in\mathcal{L}^{2}$ (the square integrable function space), then $\tilde{w}(k)\in\mathcal{L}^{2}$, since the denominator is bounded away from 0 and well behaved at infinity. Hence, $w(x)\in\mathcal{L}^{2}$, the operator $(\tilde{\mathbb{V}}+m)$ is invertible and $-m$ does not belong to the spectrum. If $-m\in[-M,+\infty)$, $\tilde{w}(k)$ has isolated poles and for generic $G\in\mathcal{L}^{2}$ it does not belong to $\mathcal{L}^{2}$, neither does $w(x)$. The resolvent is unbounded and, $-m$ belongs to the continuous spectrum of $\mathbb{V}$.


We conclude that we mus impose $\tilde{\mathbb{V}}(k^{2})\geq0$ to have $\mathbb{V}\geq 0$. Moreover, since $\mathbb{V}$ has a continuous spectrum without eigenvalues in $\mathcal{L}^{2}$, $H_{\text{RED}}$
with domain in $\mathcal{L}^{2}$ satisfies $H_{\text{RED}}>0$ if and only if $\tilde{\mathbb{V}}(k^{2})\geq0$.

We now determine the restrictions on the coupling parameters to have $\tilde{\mathbb{V}}(k^{2})\geq0$. We get  \begin{equation}\label{eq98}
     \tilde{\mathbb{V}}(u)=u\left(|\hat{\beta}_{5}|u^{3}-\hat{\beta}_{3}u^{2}-\hat{\beta}_{1}u+\beta\right),  
    \end{equation}
where $u=k\cdot k\geq0$, and {$\beta>0$}, {$\hat{\beta}_{5}<0$} from previous assumptions. 
\begin{enumerate}
    \item If $\hat{\beta}_{1}\leq 0$ and $\hat{\beta}_{3}\leq 0$ then $\tilde{\mathbb{V}}(k^{2})\geq0$.
    \item If $\hat{\beta}_{1}>0$, the cubic polynomial in the right hand side of (\ref{eq98}) may have: i) 2 positive and 1 negative real roots or ii) 2 complex conjugate roots and 1 negative real root. In order to have $\tilde{\mathbb{V}}(u)\geq0$ we must have case ii) which arises by imposing the discriminant $\Delta$ of the cubic equation to be negative 
    \begin{equation}\label{eq99}
    \Delta=\left(18|\hat{\beta}_{5}|\hat{\beta}_{1}+4\hat{\beta}^{2}_{3}\right)\hat{\beta}_{3}\beta+\left(\hat{\beta}^{2}_{3}+4|\hat{\beta}_{5}|\hat{\beta}_{1}\right)\hat{\beta}^{2}_{1}-27|\hat{\beta}_{5}|^{2}\beta^{2}<0.  
    \end{equation}
    Hence, if $\hat{\beta}_{1}>0$ then $\tilde{\mathbb{V}}(k^{2})\geq0$ if and only if $\Delta<0$.
    \item If $\hat{\beta}_{1}\leq 0$ and $\hat{\beta}_{3}> 0$, the cubic polynomial may have the same i) or ii) cases as before. Consequently, if $\hat{\beta}_{1}\leq 0$ and $\hat{\beta}_{3}> 0$ then $\tilde{\mathbb{V}}(k^{2})\geq0$ if and only if $\Delta<0$.
\end{enumerate}
We conclude that the ground state is stable if:
 $\beta>0$, $\hat{\beta}_{5}<0$ (from previous assumptions) and $\hat{\beta}_{1}>0$, $\Delta<0$, or $\hat{\beta}_{1}\leq 0$, $\hat{\beta}_{3}\leq 0$, or $\hat{\beta}_{1}\leq 0$, $\hat{\beta}_{3}> 0$, $\Delta<0$.

 {The previous result, come from the fact that the cubic polynomial $\mathbb{P}(u)\equiv |\hat{\beta}_{5}|u^{3}-\hat{\beta}_{3}u^{2}-\hat{\beta}_{1}u+\beta$ in the right hand side of Eq. (\ref{eq98}), always contains (at least) one negative root, because $\mathbb{P}( -\infty) = -\infty$ and $\mathbb{P}(0) = \beta > 0$ (another way to check it, is by using the Descartes's sign rule).
Moreover, here we will not consider in the discussion the case of multiple roots, because it implies fine tuning on the parameters, a relation between the  parameters $\hat{\beta}_{1}$, $\hat{\beta}_{3}$ and $\hat{\beta}_{5}$. In this case, one needs to demand the discriminant $\Delta = 0$. }
 

\section{Renormalizability by power counting}\label{sec4}

\subsection{The propagation of the gravity--gauge vector H--L model}

It follows from the analysis of the evolution of the physical modes, that is the transverse--traceless tensorial modes and the transverse vectorial modes, the propagators of the anisotropic H--L model are 
\begin{equation}\label{eq107}
\langle h^{TT}_{ij}h^{TT}_{kl} \rangle=  \frac{\mathbb{P}_{ijkl}^{TT}}{\frac{\omega^{2}}{2\kappa}-\beta\vec{k}^{2}+\hat{\beta}_{1}\vec{k}^{4}+\hat{\beta}_{3}\vec{k}^{6}+\hat{\beta}_{5}\vec{k}^{8}},
\end{equation}
while for the vector--gauge field sector is expressed by
\begin{equation}\label{eq108}
\langle \xi^{T}_{i}\xi^{T}_{j} \rangle=  \frac{\theta_{ij}}{\frac{\omega^{2}}{2\kappa}-\beta\vec{k}^{2}+\hat{\beta}_{1}\vec{k}^{4}+\hat{\beta}_{3}\vec{k}^{6}+\hat{\beta}_{5}\vec{k}^{8}},
\end{equation}
where 
\begin{equation}\label{eq109}
\mathbb{P}^{TT}_{ijkl}=\frac{1}{2}\left(\theta_{ik}\theta_{jl}+\theta_{il}\theta_{jk}-\theta_{ij}\theta_{kl}\right), \quad \theta_{ij}=\delta_{ij}-\frac{k_{i}k_{j}}{\vec{k}^{2}}.    
\end{equation}
We notice that the propagator of the gravitational modes is the same as in the original Ho\v{r}ava's proposal, except that there is a new contribution $\hat{\beta}_{5}k^{8}$ inherited from the Kaluza--Klein 4+1 formulation. {Also, the denominator in (\ref{eq107}) and (\ref{eq108}) are the same, due to their 4+1 origin.  }

\subsection{Power counting}

The main point in this analysis is to determine the power counting of the derivatives acting on $h^{TT}_{ij}$ and $\xi^{T}_{i}$ when solving $h^{T}$ and $n$ from the second class constraints (\ref{eq45})--(\ref{eq46}). They can be rewritten as
\begin{equation}\label{eq75new}
    \mathbb{M}\phi=h\left(h^{TT}_{ij}; \xi^{T}_{i}; h^{T}; n\right), \quad \phi=\left(\begin{array}{c}
h^{T} \\
n
\end{array}\right).
\end{equation}
We can solve perturbatively these constraints. At first order $\phi=0$, we replace it on the right hand side and solve at second order for $\phi$. At each step, we replace the previous solution on the right hand side of (\ref{eq75new}) and solve dor $\phi$. The highest order in derivatives in $h^{T}$ is, by construction of the potential, $2z$. The main requirement is that the maximum derivative power on $\mathbb{M}$ is also $2z$. Consequently we must impose condition (\ref{eq80}) on the coupling parameters. In this way the contribution of $h^{T}$ and $n$ to the vertices does not add positive powers of momenta. Hence, the maximum powers of momenta at each vertex is $2z$ (see appendix \ref{appen}).

The power counting renormalization of the 1PI diagrams follow from the propagator (\ref{eq108}) and (\ref{eq109}), internal lines (I), loops integrals (L) and vertex (V). We consider the conventionally normalized fields \cite{weinberg}, that is, its dimensions follow from the requirement that $\hat{\beta}_{5}$ is dimensionless. Hence, for the internal lines 
\begin{equation}\label{eq113}
\langle h^{TT}_{ij}h^{TT}_{kl} \rangle_{\omega,\vec{k}}\rightarrow \Xi^{-2z}, \quad  \langle \xi^{T}_{i}\xi^{T}_{j} \rangle_{\omega,\vec{k}}\rightarrow \Xi^{-2z}
\end{equation}
and for the loop integrals 
\begin{equation}\label{eq115}
\int d\omega d^{d}k\rightarrow \Xi^{d+z}.    
\end{equation}
From a previous argument, the vertices can bring at most $2z$ powers of momenta. Then, the superficial degree of divergence satisfies
\begin{equation}\label{eq116}
D\leq \left(d+z\right)L+2z\left(V-I\right).    
\end{equation}
Using the identity $L-1=I-V$ one finally obtain 
\begin{equation}
D\leq \left(d-z\right) L+2z.    
\end{equation}
In the model under consideration $d=3$, $z=4$, then
\begin{equation}\label{deeee}
D\leq -L+8.    
\end{equation}
The anisotropic model is then power counting renormalizable. Moreover, the consistency of the Hamiltonian formulation ensures the unitarity of the theory. 

\section{Concluding Remarks}\label{sec5}

In this work, the analysis of the pure anisotropic gravity--gauge vector coupling in the non--projectable Ho\v{r}ava--Lifshitz gravity framework, at all energy scales, was performed. We showed that the theory is consistent. It propagates solely the transverse. traceless--tensorial modes together with the transverse vectorial modes as in the Einstein--Maxwell theory and it is renormalizable by power counting provided the restrictions in Section \ref{sec3} together with condition (\ref{eq80}) are satified. No ghost fields are present, the model is manifestly unitary.

We started with a 4+1--dimensional non--projectable Ho\v{r}ava--Lifshitz theory subject to a geometrical restriction (\ref{eq17new}). After that, a dimensional reduction process is performed, arriving to a 3+1--dimensional theory at the KC point, $\lambda=1/3$, (\ref{new12}), with the first class constraints (\ref{gauss1})--(\ref{mome}) and the second class constraints $P_{N}=0$, $P\equiv\gamma_{ij}{p^{ij}}=0$, (\ref{eq45}) and (\ref{eq46}). The resulting theory, is invariant under FDiff and $U(1)$ symmetry groups. In this regard, the first class constraint (\ref{gauss1}) is the generator of $U(1)$ symmetry transformation, the equivalent to Gauss's law in the Maxwell theory.
By analyzing the constraints structure we show that the theory is completely consistent and propagates only the transverse--traceless tensorial and transverse vectorial degrees of freedom. On the other hand, the constraint $P\equiv\gamma_{ij}{p^{ij}}=0$, is exactly the same as in the 3+1 dimensional  Ho\v{r}ava--Lifshitz gravity at the KC point model \cite{r18}.

To guarantee that the reduced theory is
power--counting renormalizable at the UV scale,
it is necessary to incorporate all terms up to $z=4$ order. Besides the contributions of the Riemann tensor to the potential, the ones constructed from the Weyl tensor become also relevant. As can be seen in Eqs. (\ref{eq54})--(\ref{eq57}) the marginal operators are up to order eight in spatial derivatives and the relevant deformations compromise two, four and six order spatial derivatives. To solve the non--linear second class constraints, a perturbative approach has been employed. From this analysis, the extra degree of freedom $h^{T}$ and the auxiliary field $n$, are determined in terms of the $h^{TT}_{ij}$ and $\xi^{T}$ modes. We proved the stability of the ground state provided that  some restrictions on the coupling constants, see Section \ref{sec3}, are satisfied. We determined, the propagator for each sector Eqs. (\ref{eq107}) and (\ref{eq108}), where it can be inferred the same dispersion relation at all energy scales. Finally, we showed that the theory is power--counting renormalizable, provided the restrictions obtained in Section \ref{sec3} and condition (\ref{eq80}) are satisfied.

\section*{Acknowledgements}
F. Tello-Ortiz acknowledges financial support by
project ANT–-2156  at
Universidad de Antofagasta, Chile. 

\appendix

\section{Interaction terms}\label{appen}

The main aim of this appendix, is to analyze the possible divergences in the UV regime coming from the interactions (terms beyond the quadratic order in the fields). The interaction terms compromise a pure gravitational interaction, a pure vector interaction and a mixture between them. To address this point, we shall do a qualitatively study of the structure of the interactions, being necessary to go beyond the linear order in perturbations. In turn, this requires solving the second class constraints at higher order in perturbations, since in principle the first class constraints, can be treated by the usual quantization techniques for gauge systems. 
The second class constraints  determine $h^{T}$ and $n$. We use a perturbative approach in order to solve the constraints to all orders. The main point is to show that when solved in terms of the $h^{TT}_{ij}$ modes, they do not introduce new derivatives to the interaction. To illustrate how it works in this case, we present the second class constraint ${\mathcal{H}}$ at second order in perturbations 
\begin{equation}\label{eq110}
\begin{split}
2 \epsilon\left(\mathbb{D}_{2} h^{T}+\mathbb{D}_{3} n\right)=\epsilon^{2}\bigg[-2 \kappa\, \vartheta_{i j}^{T T} \vartheta_{i j}^{T T}+\frac{\hat{\beta}_{1}}{4} \partial^{2} h_{i j}^{T T} \partial^{2} h_{i j}^{T T}+\frac{\hat{\beta}_{3}}{4} \partial^{2} \partial_{i} h_{j k}^{T T} \partial^{2} \partial_{i} h_{j k}^{T T} +\frac{\hat{\beta}_{5}}{4}\partial^{4}h^{TT}_{ij}\partial^{4}h^{TT}_{ij} & \\ +\bigg(\beta+\alpha_{1} \partial^{2}+\alpha_{3} \partial^{4}+\alpha_{5}\partial^{6}\bigg)\left( h_{i j}^{T T} \partial^{2} h_{i j}^{T T}+\frac{3}{4} \partial_{i} h_{j k}^{T T} \partial_{i} h_{j k}^{T T}-\frac{1}{2} \partial_{i} h_{j k}^{T T} \partial_{k} h_{i j}^{T T}\right) &\\
-\kappa\,\chi^{T}_{i}\chi^{T}_{i} +\kappa_{1}\bigg(\partial_{i}\partial_{j}\xi^{T}_{k}\partial_{i}\partial_{j}\xi^{T}_{k} -2\partial_{i}\partial_{j}\xi^{T}_{k}\partial_{i}\partial_{k}\xi^{T}_{j}+\partial_{i}\partial_{k}\xi^{T}_{j}\partial_{i}\partial_{k}\xi^{T}_{j}\bigg)&\\ +\kappa_{2}\bigg(\partial_{m}\partial_{i}\partial_{j}\xi^{T}_{k}\partial_{m}\partial_{i}\partial_{j}\xi^{T}_{k}-2\partial_{m}\partial_{i}\partial_{j}\xi^{T}_{k}\partial_{m}\partial_{i}\partial_{k}\xi^{T}_{j}+\partial_{m}\partial_{i}\partial_{k}\xi^{T}_{j}\partial_{m}\partial_{i}\partial_{k}\xi^{T}_{j}\bigg)&\\+\kappa_{3}\bigg(\partial^{2}\partial_{i}\partial_{j}\xi^{T}_{k}\partial^{2}\partial_{i}\partial_{j}\xi^{T}_{k}-2\partial^{2}\partial_{i}\partial_{j}\xi^{T}_{k}\partial^{2}\partial_{i}\partial_{k}\xi^{T}_{j}+\partial^{2}\partial_{i}\partial_{k}\xi^{T}_{j}\partial^{2}\partial_{i}\partial_{k}\xi^{T}_{j}\bigg) &\\
+\hat{\hat{\beta}}_{1}\partial^{2}\xi^{T}_{i}\partial^{2}\xi^{T}_{i}+\hat{\hat{\beta}}_{3}\partial_{k}\partial^{2}\xi^{T}_{i}\partial_{k}\partial^{2}\xi^{T}_{i}+\hat{\hat{\beta}}_{5}\partial^{4}\xi^{T}_{i}\partial^{4}\xi^{T}_{i}\bigg],
\end{split}
\end{equation}
where 
\begin{equation}\label{eq111}
\hat{\hat{\beta}}_{1}\equiv \frac{\beta_{1}}{2}+\kappa_{1}, \quad \hat{\hat{\beta}}_{3}\equiv \frac{\beta_{3}}{2}+\kappa_{2} , \quad \hat{\hat{\beta}}_{5}\equiv \frac{\beta_{5}}{2}+\kappa_{3},   
\end{equation}
and the operators $\mathbb{D}_{2}$ and $\mathbb{D}_{3}$ were defined in Eq. (\ref{eq75}). In obtaining the above result we have replaced the solution 
of $h^{T}$ and $n$ at first order \i.e., $h^{T}=n=0$ in all second order terms weighted by $\epsilon^{2}$. To obtain the solution at the next orders, the corresponding previous order solutions must be substituted on the right hand member. Hence, the second class constraints ${\mathcal{H}}$ and ${\mathcal{C}}$ yields linear equations for the variables $h^{T}$ and $n$ at any order in perturbations where the operator acting on them is just the matrix operator $\mathbb{M}$. 

In the present situation it is enough to know the distribution of the momenta at the UV regime. Then, one can consider only those terms that contribute with the highest power of momenta in the Fourier space. At any order in perturbations the highest number of spatial derivatives contained in both $\bar{\mathcal{H}}$ and $\bar{\mathcal{C}}$ is exactly the same number of derivatives acting on the fields $h^{T}$ and $n$. As can be appreciated from Eq. (\ref{eq110}) the maximum order in spatial derivatives acting on $h^{T}$ and $n$ is eight while the total number of derivatives on the right hand member acting on products of $h^{TT}_{ij}$ and $\xi^{T}_{i}$ is also eight. Moreover, the constraints $\bar{\mathcal{H}}$ and $\bar{\mathcal{C}}$ do not contain spatial derivatives of the conjugate momenta. Thereby, for the second and higher order in perturbations, one can model (schematically) the dominant part of the solutions at the UV fixed point as follows 
\begin{equation}\label{eq112}
\begin{split}
h^{T}, n \sim\left(\frac{1}{\left(\partial_{m}\right)^{2 z}}\left(\partial_{n}\right)^{2 z}\right)\left(h_{i j}^{T T} \cdots h_{k l}^{T T}\right),\quad \frac{1}{\left(\partial_{m}\right)^{2 z}}\left(h_{i j}^{T T} \cdots h_{k l}^{T T} \vartheta_{p q}^{T T} \vartheta_{r s}^{T T}\right), \quad \left(\frac{1}{\left(\partial_{m}\right)^{2 z}}\left(\partial_{n}\right)^{2 z}\right)\left(\xi^{T}_{i } \cdots \xi^{T}_{j}\right), & \\ \frac{1}{\left(\partial_{m}\right)^{2 z}}\left(\xi_{i}^{T} \cdots \xi_{k }^{T} \chi_{q}^{T} \chi_{p}^{ T}\right), \quad \left(\frac{1}{\left(\partial_{m}\right)^{2 z}}\left(\partial_{n}\right)^{2 z}\right)\left(h_{i j}^{T T} \cdots \xi_{k}^{T}\right).
\end{split}
\end{equation}
The above entails that at the highest order in derivatives, the operator $\mathbb{M}$ can be expressed as the differential operator $\partial^{2z}$ times a matrix of dimensionless coupling constant with determinant (\ref{eq80}) which must be different from zero. This condition is a fundamental one, otherwise the number of derivatives in the denominator of (\ref{eq112}) would be less than $2z$. In that case the contribution  of $h^{T}$ and $n$ to the interaction terms will have positive powers of momenta and could imply the non--renormalizability of the theory. 

The above argument shows that under the condition (\ref{eq80}) the contribution of $h^{T}$ and $n$ to the interactions, in terms of $h^{TT}_{ij}$ and $\xi^{T}_{i}$,  does not change number of derivatives of the interaction terms. From the schematic solution provided by Eq. (\ref{eq112}) one can see (for example) that the cubic interactions at $2z$ order like $h^{T}h^{TT}_{ij}\partial^{8}h^{TT}_{ij}$ and $h^{T}\xi^{T}_{i}\partial^{8}\xi^{T}_{i}$, after substituting the solution for $h^{T}$, keep the vertex contribution with 8 powers of momenta. We can now calculate the superficial degree of divergence of the theory, determining if under the previous considerations the full theory is power--counting renormalizable or not.


\section*{References}
\begin{thebibliography}{99}


\bibitem{r1}  P. Ho\v{r}ava, \emph{Phys. Rev. D} \textbf{79}, 084008 (2009).

\bibitem{lif} E. M. Lifshitz, \emph{Zh. Eksp. Teor. Fiz.} \textbf{11}, 255 (1941).

\bibitem{r3}  D. Blas, O. Pujolas and S. Sibiryakov, \emph{Phys. Rev. Lett.} \textbf{104}, 181302 (2010).


\bibitem{r2} R. Arnowitt, S. Deser and C. Misner, \emph{Gen. Relativ. Gravit.} \textbf{40}, 1997 (2008).

\bibitem{r6} J. Bellor\'{i}n, A. Restuccia and A. Sotomayor, \emph{Phys. Rev. D} \textbf{87}, 084020 (2013).

\bibitem{r19}J. Bellorín, and A. Restuccia,
\emph{Int. J. Mod. Phys. D} \textbf{21}, 1250029 (2012). 

\bibitem{r7}  K. S. Stelle, \emph{Phys.
Rev. D} \textbf{16}, 953 (1977). 

\bibitem{r4} C. Charmousis, G. Niz, A. Padilla and P. M. Saffin, \emph{JHEP} \textbf{0908}, 070 (2009).

\bibitem{r5}
D.Blas, O.Pujolas and S.Sibiryakov,
\emph{Phys. Lett. B} \textbf{688}, 350 (2010).

\bibitem{rr5} A. Papazoglou, T. P. Sotiriou, \emph{Phys. Lett. B} \textbf{685}, 197 (2010).

\bibitem{bara} E. Barausse, \emph{Phys. Rev. D} \textbf{100}, 084053 (2019).

\bibitem{r8}  P. Ho\v{r}ava and C. M. Melby--Thompson, \emph{Phys. Rev. D} \textbf{82}, 064027  (2010). 

\bibitem{l1} G. Cognola, R. Myrzakulov, L. Sebastiani, S. Vagnozzi and S. Zerbini,
  \emph{Class. Quant. Grav.}
   \textbf{33}, 225014 (2016).

\bibitem{l2} A. Casalino, M. Rinaldi, L. Sebastiani and S. Vagnozzi,
  \emph{Class. Quant. Grav.}
   \textbf{3}, 017001 (2019).

\bibitem{r24444} J. Bellorín and A. Restuccia, \emph{Int. J. Mod. Phys. D} \textbf{27},  1750174 (2018).

\bibitem{r9}  T. Jacobson and D. Mattingly, \emph{Phys.
Rev. D} \textbf{64}, 024028 (2001).

\bibitem{r10}  T. Jacobson, \emph{Phys. Rev.
D} \textbf{81}, 101502  (2010).

\bibitem{r11} T. Jacobson, \emph{Phys.
Rev. D} \textbf{89}, 081501 (2014).

\bibitem{r12} J. Hartong and N. A. Obers, \emph{JHEP} \textbf{1507}, 155 (2015).

\bibitem{r23}
  S.Shin and M.I.Park,
  \emph{JCAP} \textbf{1712}, 033 (2017).

\bibitem{r24} G. Leon and A. Paliathanasis,  \emph{Eur. Phys. J. C} \textbf{79}, 746 (2019).

\bibitem{r25} M. A. Zadeh, arXiv: 1912.06495 [gr-qc] (2020).

{\bibitem{r26} F. Gao and J. Llibre, \emph{Universe} \textbf{7}, 445 (2021).}

\bibitem{r27} R. Bluhm, H. Bossi and y Wen \emph{Phys. Rev. D} \textbf{100}, 084022 (2019).

\bibitem{r28} M. Gomes, T. Mariz, J. R. Nascimento, A. yu. Petrov and A. J. da Silva, \emph{Eur. Phys. J. C} \textbf{80}, 518 (2020).


\bibitem{r13} J. Kluson, \emph{JHEP} \textbf{1007}, 038 (2010).

\bibitem{r14} W. Donnelly
and T. Jacobson, \emph{Phys. Rev. D} \textbf{84},  104019 (2011).

\bibitem{r16} J. Bellor\'{i}n and A. Restuccia, \emph{Phys. Rev. D} \textbf{84}, 104037 (2011).

\bibitem{r15} J. Bellorín, A. Restuccia and A. Sotomayor, \emph{Phys.
Rev. D} \textbf{85}, 124060 (2012).

\bibitem{new1} J. Kluson, \emph{Phys. Rev D} \textbf{83}, 044049 (2011).

\bibitem{new2} J. Kluson, \emph{Phys. Rev D} \textbf{82}, 044004 (2010).

\bibitem{new3} S. Koh and S. Shin, \emph{Phys. Lett. B} \textbf{696}, 426 (2011).

\bibitem{new4} M. Chaichian, J. Kluson and M. Oksanen, \emph{Phys. Rev. D} \textbf{92}, 104043 (2015).

\bibitem{r18} J. Bellorín and A. Restuccia, \emph{Phys.
Rev. D} \textbf{94}, 064041 (2016).

\bibitem{blas2} A. O. Barvinsky, D. Blas, M. Herrero--Valea, S. M. Sibiryakov and C. F. Steinwachs,
\emph{Phys. Rev. D} \textbf{93}, 064022 (2016).

\bibitem{blas3} A. O. Barvinsky, M. Herrero--Valea, and S. M. Sibiryakov, \emph{Phys. Rev. D} \textbf{100}, 026012 (2019).

\bibitem{barvinsky} Andrei O. Barvinsky,  D. Blas, M. Herrero-Valea, S. M. Sibiryakov and C. F. Steinwachs, \emph{Phys. Rev. Lett.} \textbf{119}, 211301 (2017).
  
\bibitem{by} J. Bellorín and B. Droguett, \emph{Phys. Rev. D} \textbf{101}, 084061 (2020).  

\bibitem{chin1} Fu-Wen Shu and T. Zhang, \emph{Symmetry} \textbf{13} (1), 100 (2021). 

\bibitem{r32}
  G. t'Hooft and M. J. G. Veltman,
  \emph{Ann. Inst. H. Poincare Phys. Theor. A} \textbf{20}, 69 (1974).

\bibitem{r31}  M. Pospelov and y. Shang,
\emph{Phys. Rev. D} \textbf{85}, 105001 (2012).

\bibitem{ian} I. Kimpton and A. Padilla\emph{J. High Energ. Phys.} \textbf{2013}, 133 (2013).

\bibitem{r35} M. Colombo, A. E. G\"{u}mr\"{u}k\c{c}uo\u{g}lu and T. P. Sotiriou, \emph{Phys. Rev. D} \textbf{91}, 044021 (2015).

\bibitem{r36} M. Colombo, A. E. G\"{u}mr\"{u}k\c{c}uo\u{g}lu and T. P. Sotiriou, \emph{Phys. Rev. D} \textbf{92}, 064037 (2015).

\bibitem{r29} J. Bellor\'{i}n, A. Restuccia and F. Tello--Ortiz, \emph{Phys. Rev. D} \textbf{98}, 104018 (2018).

\bibitem{r30} A. Restuccia and F. Tello--ortiz \emph{Eur. Phys. J. C} \textbf{80}, 86 (2020).

\bibitem{wh1} A. Restuccia and F. Tello-Ortiz, \emph{Eur. Phys. J. C} \textbf{81}, 447 (2021).


\bibitem{wh2} J. Bellorín, A. Restuccia and A. Sotomayor, \emph{Phys. Rev. D} \textbf{90}, 044009 (2014).

\bibitem{will} C. M. Will, \emph{Living Rev. Relativ.} \textbf{9}, 3 (2005).

\bibitem{feno1} B. P. Abbott et al., \emph{Astrophys. J.} \textbf{848}, L12 (2017).

\bibitem{feno2} B. P. Abbott et al. (LIGO Scientific
Collaboration and Virgo Collaboration),
\emph{Phys. Rev. Lett.} \textbf{119}, 161101 (2017).

\bibitem{feno3} B. P. Abbott et al. (Virgo, Fermi-GBM, INTE-
GRAL, LIGO Scientific), \emph{Astrophys. J.} \textbf{848}, L13 (2017).

\bibitem{feno5} A. E. Gumrukcuoglu, M. Saravani and
T. P. Sotiriou, \emph{Phys. Rev. D} \textbf{97}, 024032 (2018)

\bibitem{feno6} O. Ramos and E. Barausse,
\emph{Phys. Rev. D} \textbf{99}, 024034 (2019).

\bibitem{feno7} T. Zhang, F. Shu, Q. Tang and D. Du, \emph{Eur. Phys. J. C} \textbf{80}, 1062 (2020). 



\bibitem{r37} M. Visser, \emph{Phys.
Rev. D} \textbf{80}, 025011 (2009).

\bibitem{r38} M. Visser, (2009) arXiv:0912.4757 [hep-th].

\bibitem{an} D. Anselmi and M. Halat, \emph{Phys. Rev. D} \textbf{76}, 125011 (2007).

\bibitem{an1} D. Anselmi, \emph{JHEP} \textbf{0802}, 051 (2008).

\bibitem{an2} D. Anselmi, \emph{Annals Phys.} \textbf{324}, 874 (2009).

\bibitem{weinberg} S. Weinberg, \emph{The Quantum Theory of Fields, Vol. I}, (USA: Cambridge University Press), (1995).

\bibitem{r40} T. Regge and C. Teitelboim, \emph{Ann. Phys.} \textbf{88}, 286 (1974).


\bibitem{r42} D. Grumiller and R. Jackiw , \emph{Int. J. Mod. Phys. D} \textbf{15}, 2075 (2006).

\bibitem{r43} R. Jackiw, \emph{SIGMA} \textbf{3}, 091 (2007).

\bibitem{r44} D. Grumiller and R. Jackiw , \emph{Phys. Lett. A} \textbf{372}, 2547 (2008).

\end{thebibliography}
\end{document}